\documentstyle[12pt,aasms4,epsfig]{article}

\newcommand{\etal} {{\it et~al.\ }}
\def\RA#1 #2 #3 #4 { $#1^h\,#2^m\,#3^s.#4 $}
\def\ra#1 #2 #3 { $#1^h\,#2^m\,#3^s. $}
\def\dec#1 #2 #3 #4{$#1^\circ\,#2'\,#3^{\prime \prime}.#4$}
\def\erra#1 #2 {$#1^s.#2 $}
\def\errd#1 #2 {$#1^{\prime \prime}.#2 $}

\makeatletter

\let\@internalcite\cite
\def\cite{\def\@citeseppen{-1000}%
    \def\@cite##1##2{(##1\if@tempswa , ##2\fi)}%
    \def\citeauthoryear##1##2##3{##1 ##3}\@internalcite}

\def\citeNP{\def\@citeseppen{-1000}%
    \def\@cite##1##2{##1\if@tempswa , ##2\fi}%
    \def\citeauthoryear##1##2##3{##1 ##3}\@internalcite}
\def\citeN{\def\@citeseppen{-1000}%
    \def\@cite##1##2{##1\if@tempswa , ##2)\else{)}\fi}%
    \def\citeauthoryear##1##2##3{##1 (##3}\@citedata}
\def\citeA{\def\@citeseppen{-1000}%
    \def\@cite##1##2{(##1\if@tempswa , ##2\fi)}%
    \def\citeauthoryear##1##2##3{##1}\@internalcite}
\def\citeANP{\def\@citeseppen{-1000}%
    \def\@cite##1##2{##1\if@tempswa , ##2\fi}%
    \def\citeauthoryear##1##2##3{##1}\@internalcite}
\def\shortcite{\def\@citeseppen{-1000}%
    \def\@cite##1##2{(##1\if@tempswa , ##2\fi)}%
    \def\citeauthoryear##1##2##3{##2 ##3}\@internalcite}
\def\shortciteNP{\def\@citeseppen{-1000}%
    \def\@cite##1##2{##1\if@tempswa , ##2\fi}%
    \def\citeauthoryear##1##2##3{##2 ##3}\@internalcite}
\def\shortciteN{\def\@citeseppen{-1000}%
    \def\@cite##1##2{##1\if@tempswa , ##2)\else{)}\fi}%
    \def\citeauthoryear##1##2##3{##2 (##3}\@citedata}
\def\shortciteA{\def\@citeseppen{-1000}%
    \def\@cite##1##2{(##1\if@tempswa , ##2\fi)}%
    \def\citeauthoryear##1##2##3{##2}\@internalcite}
\def\shortciteANP{\def\@citeseppen{-1000}%
    \def\@cite##1##2{##1\if@tempswa , ##2\fi}%
    \def\citeauthoryear##1##2##3{##2}\@internalcite}
\def\citeyear{\def\@citeseppen{-1000}%
    \def\@cite##1##2{(##1\if@tempswa , ##2\fi)}%
    \def\citeauthoryear##1##2##3{##3}\@citedata}
\def\citeyearNP{\def\@citeseppen{-1000}%
    \def\@cite##1##2{##1\if@tempswa , ##2\fi}%
    \def\citeauthoryear##1##2##3{##3}\@citedata}

\def\@citedata{%
	\@ifnextchar [{\@tempswatrue\@citedatax}%
				  {\@tempswafalse\@citedatax[]}%
}

\def\@citedatax[#1]#2{%
\if@filesw\immediate\write\@auxout{\string\citation{#2}}\fi%
  \def\@citea{}\@cite{\@for\@citeb:=#2\do%
    {\@citea\def\@citea{, }\@ifundefined
       {b@\@citeb}{{\bf ?}%
       \@warning{Citation `\@citeb' on page \thepage \space undefined}}%
{\csname b@\@citeb\endcsname}}}{#1}}%

\def\@citex[#1]#2{%
\if@filesw\immediate\write\@auxout{\string\citation{#2}}\fi%
  \def\@citea{}\@cite{\@for\@citeb:=#2\do%
    {\@citea\def\@citea{; }\@ifundefined
       {b@\@citeb}{{\bf ?}%
       \@warning{Citation `\@citeb' on page \thepage \space undefined}}%
{\csname b@\@citeb\endcsname}}}{#1}}%

\def\@biblabel#1{}

\newlength{\bibhang}
\makeatother

\begin{document}

\title{Deep Radio Imaging of Globular Clusters and the Cluster Pulsar Population}

\author{A. S. Fruchter}
\affil{Space Telescope Science Institute, 3700 San Martin Drive,
       Baltimore, MD 21210}

\author{W. M. Goss}
\affil{NRAO,P.O. Box 0,Socorro, NM 87801}

\begin{abstract}
We have obtained deep multifrequency radio observations
of  seven
globular clusters using the Very Large Array
and the Australia Telescope Compact Array.  Five of these,
NGC~6440, NGC~6539, NGC~6544, NGC~6624 and Terzan~5 had 
previously 
been detected in a shallower survey for steep spectrum
radio sources in globular clusters \cite{fg90}. 
The sixth,
the rich globular cluster, Liller 1, 
had heretofore been undetected in the radio, and the 
seventh, 47 Tucanae, was not included in our original
survey.
High resolution 6 and 20 cm images of three of the clusters,
NGC~6440, NGC~6539, NGC~6624
reveal only point sources coincident with pulsars which
have been discovered subsequent to 
our first imaging survey.
21 and 18 cm images reveal several point sources within
a few core-radii of the center of 47 Tuc.  Two of these
are identified pulsars, and a third, which is both variable
and  has a steep spectrum, is also most likely a pulsar
previously identified by a pulsed survey.
However, the 6, 20 and 90 cm images of NGC~6544, Liller~1 and Terzan~5 
display strong steep-spectrum emission which cannot be associated with
known pulsars. 
The 90 cm flux densities
of Terzan~5, Liller~1 and NGC 6544 are 35, 9 and 6 mJy, respectively, implying
that a number of bright pulsars in these clusters have been
hidden from pulsed searches.  

The image of the rich cluster Terzan 5
displays numerous point sources within $30''$, or 4 core radii of
the cluster center.  The density of these objects rises rapidly toward
the core, where an elongated region of emission is found.
This feature may be due to the presence of a higher density of point sources
than can be resolved by the $1''$ resolution of our data.  
The brightest individual sources,
as well as the extended emission, possess the steep spectra expected of 
pulsars.   Furthermore, the flux distribution of the sources 
agrees well with the standard pulsar luminosity function.
The total luminosity and number of
objects observed suggest that Terzan~5 contains
more pulsars than any other Galactic globular cluster.
\end{abstract}

\section{Introduction}

Low mass x-ray binaries (LMXBs) announce themselves from across the galaxy
to telescopes with effective areas measured in square centimeters, but
some of their descendants, binary millisecond pulsars (MSPs) are far less
vocal.  The majority of MSPs are difficult for even the largest radio
telescopes to detect at a distance of a few kiloparsecs,  and the pulses
from even the bright MSPs can be obscured by the dispersion of the
interstellar medium or by doppler smearing resulting from the gravitational
tug of a binary companion.  These latter difficulties, however, can be
overcome by radio interferometric imaging.   The flux density measured
in a radio image is not affected by dispersion or doppler broadening.
Furthermore, one can generally distinguish between pulsars and background
sources by the steep spectra of pulsars and their occasional polarization.
Indeed, the first pulsar found in a globular cluster was located in
a Very Large Array (VLA) image \cite{hhb85}.

As discussed in Fruchter and Goss (1990), henceforth FG90, interferometric
imaging has another, subtle, but equally important advantage.
If the synthesized beam  of the interferometer is comparable in diameter
to the core of the cluster, then the flux density one observes will frequently
be the sum of the emission from many pulsars in the cluster.  If the
number of pulsars in  a clusters is large, this sum can substantially
exceed the flux density from the single brightest pulsar in the cluster.
The luminosity function of pulsars, as measured in the field, 
has the form
\begin{equation}
\frac{dn}{dl} = \frac{1}{L^p} \, ,
\end{equation}
where $L$, the pulsar luminosity,
is bounded below by a minimum luminosity, $L_{min}$,
of order $\sim 0.3$ mJy kpc$^2$
at 20 cm (1400 MHz), and the
power, $p$, is approximately 2 \cite{mt77,dtws85,sto87a}.   
Limited observational
evidence has so far suggested that the luminosity function
of pulsars in globular clusters agrees with
that of the field \cite{and92}. 
Although with this luminosity law
the expected luminosity of a pulsar (and thus a cluster)
diverges logarithmically,
it can be shown, 
that for
a given observed cluster luminosity, if $L_{tot} \gg L_{min}$,
 the expected number of pulsars in the  cluster beaming towards earth is
\begin{equation}
N_{puls} = \frac{L_{tot}}{L_{min}\ln{\left( \frac{L_{tot}}{L_{min}}\right) } }.
\end{equation}
Thus, by observing a cluster with a large beam, we not only gain the
advantage of the increased flux density, but also obtain an
estimator of the number of pulsars in the cluster.

In FG90 we described a search of 17 globular clusters for radio emission
at 20 cm.  Resolutions comparable to the core diameters were employed
to maximize the sensitivity to the integrated emission from a population
of pulsars in the clusters.  New sources were detected in 
NGC~6440, NGC~6539, NGC~6544 and Terzan~5  and we confirmed the existence of
a controversial source in NGC~6624.  As reported in FG90,
repeat observations of these
sources at 20 cm, and observations of many of them at 6 cm, suggested
that with the possible exception of the source in NGC~6624 all of the
sources were due to radiation from pulsars, as a result of their observed
variability or steep spectrum.  Later, pulsed searches, from Parkes
and Jodrell revealed individual pulsars with flux densities comparable
to those we observed in three of the clusters, NGC~6440, NGC~6539 and 
NGC~6624.  Additionally, a source visible on our image of Terzan~5,
but well outside of the cluster core, was found to be an eclipsing
millisecond pulsar.

In order to better study the pulsar population in these clusters
we have reobserved these clusters with the VLA but using a substantially
smaller beam size.  Observations were taken at 20 and 6 cm, and
particular attention was paid to the two optically richest clusters in our
survey Terzan~5 and NGC~6440.   Further observations were
taken at 90 cm  of NGC~6544 and Terzan~5
in order to better characterize
the emission from clusters where emission was seen but no pulsar
found.  Finally, we have expanded our study to include two other
rich clusters, Liller~1 and 47~Tuc, which had substantial unpublished
data in the archives of the VLA and Australia Telescope Compact Array (ATCA), respectively.   Early versions of some of the results reported here
were presented in  Fruchter and Goss (1991) and Fruchter and Goss (1995).
\nocite{fg91,fg95}

\section{Observations and Results}

\subsection{Terzan 5}

The globular cluster Terzan~5 was observed at 20 cm in the A, BnA,
B and CnB configurations of the Very Large Array (VLA)\footnote{The A 
configuration of the VLA is its largest.  Each successive array (B,C,D) 
is approximately a factor of three smaller than its predecessor.  
Hybrid configurations ({\it e.g.} BnA),
have the northern
arm in the larger configuration, to provide a more
circular beam for southern sources.}.  In each
case, approximately four hours of data were obtained. 
All observations were done
in continuum mode, using two  50 MHz bandpasses,  centered on the
sky frequencies of 1465 and 1515 MHz.   Both right and left circular
polarizations were observed.   Terzan~5 was also observed at 6 cm
in the B and C arrays.  In both cases the observations were done
in continuum mode, employing two orthogonal polarizations in each
of two 50 MHz bandpasses, centered at the sky frequencies of 4835 and
4885 MHz.  
Here we discuss the 6~cm C array data as its beamsize ($\sim 4''$)
closely approximates that which we used to measure the extended emission
observed at 20 cm.   About six hours of data were taken on source 
in the C array during two
four hour observing sessions separated by about one week.  Finally,
Terzan~5 was reobserved on three successive days
in the A/B array.  During these observing session we rapidly switched
between 20 cm and 90 cm. 
The 20 cm observations were done as described
above.  At 90 cm we observed across two 3.125 MHz bandpasses centered
at 327.5 and 333 MHz.  Both linear polarizations were observed.  These
dual frequency
observations were designed to allow us to study both the emission from 
the center of the cluster as well as the
the low-frequency eclipse behavior of
the 11~ms pulsar 1744-24A. 

EDITOR: INSERT FIGURES 1-4 HERE

EDITOR: INSERT TABLE 1 HERE

Figure~1 shows the contours of the 20 cm image of Terzan~5 superposed
on an I band image
of the cluster taken by Taft Armandroff with the Cerro Tololo Inter-American
Observatory (CTIO) 1-m telescope.
The VLA image was made
using uniform weighting and
a 40~kilolambda taper.   This weighting produces a small beam size
($2\farcs9 \times 2\farcs9 $) which sharpens the details of the image
with the loss of some sensitivity to extended structure.
The bright radio source
about $30''$ to the west of the cluster is the eclipsing pulsar, PSR~1744-24A
\cite{lmd+90,nttf90,nt92}.  
The ``sidelobes'' and apparent elongation of this 
source are due to its time variability caused primarily
by the obscuration of the pulsar by a wind off its companion. 
Two other points sources, a northern and southern source can be seen
on either side of the central cluster emission (the positions and
flux densities {\it per beam} of the sources in Figure~1 are 
displayed in Table~1).  
Perhaps the most striking feature of the radio image, however, is the diffuse
radiation centered on the optical cluster, which can be better
viewed in the image shown in Figure~2, which has a 
beam of $5\farcs9 \times 4\farcs1$.  This image was created
using a ``robust weight'' of $R=1$ \cite{brig95},  this is a weighting
roughly intermediate between uniform weighting, which emphasizes resolution,
and natural weighting, which emphasizes sensitivity.

The flux density of the
diffuse component
is $\sim 2$~mJy at 20 cm:
stronger than the two prominent point sources in the image.
Even when only the highest resolution
data is retained and an image with a $1''$ beam is produced (with
a consequent loss of signal-to-noise) the center of the cluster
remains a continuous, elongated resolved source.  If, as we suspect,
the source of this emission is pulsars, their central density is too
large for them to be individually distinguished in our data.

The alignment of the radio and optical images is based on the
astrometric solutions for a number of
stars in the I-band image which are also found
in the HST Guide Star Catalog \cite{rlb+90}, which has
an astrometric accuracy somewhat better than $1''$.
As a number of stars in that catalog were in the I-band image, and as the
positions
the peaks of the diffuse radio and optical flux densities are indistinguishable
with this astrometry, we believe we have correctly identified the center of
the Terzan~5 cluster.  The position derived for the center of
the diffuse emission, $\alpha_{1950} = 17^h 45^m 0\fs4$, 
$\delta_{1950} = -24\arcdeg 45\arcmin 45\farcs9$, 
agrees well with that measured  by the discoverer of
the cluster, Terzan (1971);
\nocite{terzan71} however, this position
disagrees with that of Picard and Johnston (1995) \nocite{pj95}
by about 8 arcseconds. (The Picard and Johnston position is
also inconsistent with that of Terzan).
We believe
the position and association proposed here are correct.

In Figure~3 we have superposed the 6 cm images as contours
upon the 20 cm image, which is shown in greyscale.
Comparison of these images shows that 
the $\approx 1.5$~mJy 20 cm source
about ten arc seconds to the northwest of the
cluster center (labeled ``N'' in both  Table~1 and Figure~1)
possesses a spectral index, $\alpha$, of $2$, 
 and the central
diffuse emission (source C)
has an index greater than $1.4$ 
(to determine
the spectral index it is assumed that the flux density of
the object, $f(\nu)$, scales as $f(\nu) \propto \nu^{-\alpha}$
between the observed frequencies).
The only astrophysical sources likely to possess spectra this
steep are pulsars.  Indeed, it is worth noting that the spectrum
of PSR~1744-24A it is completely undetected
in the 6 cm image.

The 6 cm image also reveals the $\sim 200\,\mu$Jy source to the south-east
of the cluster (source S) to be highly variable (undectable in one image,
its flux density increased by at least a factor of several in the other taken
days later).
Furthermore, in the 20 cm BnA configuration image
obtained simultaneously with our 90 cm data
 (but which is not included in the 20 cm image shown here),
 this source has brightened
considerably to become a $\sim 1$~mJy source.  
While the rapid 6 cm variability can be ascribed to diffractive
scintillation, the cause of the long-term 20 cm variation is
less clear.   The expected diffractive scintillation
bandwidth at 20 cm 
(assuming a dispersion measure of
$\sim 200$~pc~cm$^{-3}$) 
is of order 1 MHz \cite{cwb85},
significantly smaller than the
100 MHz bandpass of the observations, and refactive scintillation
is usually limited to changes in flux density of a factor of 2 \cite{ks92}.
Scattering by caustics in the interstellar medium, 
while relatively rare, can cause such strong 
variability \cite{grbn87}.
Nonetheless, there can be
little doubt that the object is a pulsar -- the simultaneous
90 and 20 cm observations imply a spectral index of $\sim 1.75$,
far steeper than typical background sources \cite{con84}.

Indeed, both the northern and southern point sources show up strongly in our
90 cm image. Each contributes about $\sim 13$~mJy to the total flux
density of the cluster, with another $\sim 10$~mJy apparently due to
the cluster as a whole. 
As our 90 cm beamsize is 
$9''$ across, the division of the flux density into these components
must be regarded as approximate.
Unfortunately, a beamsize this large was required in order to make
the necessary 3-D imaging of the 90 cm data 
computationally feasible.  In Figure~4, we display the 90 cm
radio contours superposed on a greyscale image of the 20 cm data
displayed in Figure~1.  The r.m.s. noise in the 90 cm image is
approximately $1.3$~mJy.  Note that the only 20 cm source not also
observed at 90 cm is the one identified pulsar in the cluster, PSR~1744-24A.
Our data provides strong evidence that this pulsar
is continuously enshrouded
by material lost from the companion.  
However, there is a strong 90 cm source, which we have labelled N' (see
Figure~4 and Table~1)
 which
is not visible at all in the 20 cm image.  This source must either
have a spectral index $\alpha > 2.5$ between 90 and 20 cm, or vary
dramatically with time.   This argues strongly for the identification
of N' as a pulsar.    It is worth noting that neither N', nor any of the
other point radio sources found in Terzan 5, has a position which agrees
with either of the two known x-ray sources in that cluster \cite{jvh95};
thus, radio emission from x-ray binaries is unlikely to be 
the cause of any of the point sources reported in this cluster.

\subsection{Liller 1}

The strong central emission observed in Terzan~5 made the non-detection
\cite{jkg91}
of the nearly equally dense and equadistant cluster Liller~1 seem
particularly surprising.  While the 20 cm observations of Johnston, Kulkarni
and Goss (1991) reached the deep limit of $180\,\mu$Jy,
the data were taken  in A configuration which has a beamsize of $\sim1.5''$,
and thus  greatly over-resolve the $\sim 7''$ core radius of the
cluster \cite{dj93}.  If the cluster were to contain a number of sources,
a lower resolution image would be more sensitive.  We therefore observed
Liller 1 at 20 cm in the CnB configuration, 
which, at this wavelength, has a
 beamwidth of approximately $10''$ at the declination 
of Liller~1.  This observation revealed a 
$280 \pm 50~\mu$Jy source.    In order to determine the spectral index of
this emission we observed Liller 1 at 6 cm in the DnC configuration, and
re-reduced 90 cm B configuration archival observations made by Johnston
and Kulkarni using 3-D imaging.  We find a 90 cm (330 MHz) flux density of
$9 \pm 1$~mJy and a 6 cm (4885 MHz) flux density of $95 \pm 14$~$\mu$Jy. 

\subsection{NGC~6544}

Although NGC~6544 has been searched extensively for pulsed radio
sources \cite{bl96}
even after our discovery of a radio source in this cluster \cite{fg90},
no pulsar has been found.  Nonetheless we followed up our
low-resolution imaging of this cluster with deep 
A-array 20-cm imaging   and  B-array 6-cm observations.  Additionally
the cluster was observed by Helen Johnston and colleagues at 90 cm in
B array.  We have also reduced these archival data.   
A source of $6.5 \pm 0.8$ mJy is found at 90 cm.  At 20 cm, the flux
density is $1.2 \pm 0.1$ mJy.  No source was detected at 6 cm,
with a $3 \sigma$ upper limit to the flux density of 75 $\mu$Jy.
The 20 cm position 
$\alpha_{J2000}=$\RA 18 07 20 24 $\pm0.01$, $\, \delta_{J2000}=$
\dec -24 59 25 3 $\pm 0.1$,  
 agrees, within the rather large ($7"$) errors with the 90 cm source,
and is within $2"$ of the position found by Shawl and White (1986) \nocite{sw86}
for the optical center of the cluster.  There is no evidence
of an extended emission in any of the images.  Although the spectral index
of the source between 90 and 20 cm is only 1.1, between 90 cm and
6 cm it is greater than 2.5, strongly implying that the radio emission
from this cluster is produced by one or more pulsars.

\subsection{NGC~6440, NGC~6539, NGC~6624}

As noted previously,
after our detection of strong 20 cm emission from a number of southern
clusters, deep pulsed surveys were undertaken at Jodrell and Parkes
to locate these objects as pulsars.  In three cases, NGC~6440 \cite{mlj+89}, 
NGC~6539 \cite{dbl+93}, and NGC~6624 \cite{bbl+94} 
these surveys detected pulsars coincident with the 20 cm emission. 
All three of these clusters were
re-observed by us at 20 cm using the A-array of the VLA and at
6 cm in the B-array.  Additionally, NGC~6440, one of the
two richest clusters in our sample, was also reobserved
in the BnA, B and CnB arrays at 20 cm.  
All observations were done
in continuum mode, employing two orthogonal polarizations
in each of  two  50 MHz bandpasses,  typically centered on the
sky frequencies of 1465 and 1515 MHz at 20 cm 
4835 and 4885 MHz at 6 cm. 
NGC~6539 and NGC~6624 were each observed for  2 hours at 20 cm in the
A array and
2.2 hours at 6 cm in the B array.    NGC~6440 
was observed at 20 cm for 4.7 hours in
the A array, and 3 hours in each of the BnA, B and CnB  arrays.  
NGC~6440 was also observed at 6 cm for 3.5 hours in the B array. 
All of the resulting observations are consistent with a single
point source dominating the flux density
from each of the clusters, and each of these point sources
is consistent with the position of a known pulsar in a cluster.
In the case of NGC~6440, a small variation in flux density ($\sim 25 \%$)
was seen between images taken in the different VLA configurations; however,
the size of the variability is comparable to that
expected from the scintillation of
the known pulsar, and there is no evidence that the flux density of the
cluster increases with lower resolution, as would be the case were a 
diffuse population of pulsars contributing significantly
to the total flux density.   In Table~2 the source positions
and flux densities are compared with those of the known pulsars in 
the cluster cores.  The agreement of the positions and flux densities
is excellent.   

EDITOR: INSERT TABLE 2 HERE

\subsection{47 Tucanae}

All of the observations so far discussed were performed at the VLA.
However, the rich and relatively nearby globular cluster 47 Tuc
is too far south to be observed from New Mexico, and was therefore
not included in our original survey.  As 47 Tuc has more detected
pulsars than any other cluster, this was a particularly unfortunate
omission. 
Since FG90, however,
archival observations taken by
David McConnell and Jon Ables of 47 Tucanae
using the 6 km Australia Telescope Compact Array (ATCA), have become
public.   We therefore have reduced these data.
Two orthogonal
linear polarizations in each of two bandpasses centered on 1408 and 1708
MHz were used.  Each bandpass was divided into 4 MHz channels, and a total
effective bandwidth of 104 MHz was obtained.  The data were
taken on the 24-25 Jan 1992 (13 hours) and the 23-25 April 1992 (48 hours).
The final image is shown in Figure~5.  The image has noise of
$32 \mu$~Jy/beam, where the beam is essentially circular with a $6''$ FWHM.

Five objects with flux density greater than $200 \mu$Jy are visible
in our radio image
in a radius of about $1\farcm5$ about the center of 47 Tucanae;
their positions and flux densities are listed in Table~3.  
Based upon positional coincidence, two
of these sources can clearly be associated 
with the two of the brightest pulsars yet discovered in the cluster, 47TucC and
47TucD \cite{mlr+91,rlm+95};  furthermore, when our position for
Source 2 is used a starting position for the pulsar timing data of 47 Tuc, a
timing solution is found for the eclipsing
pulsar 47TucJ  (Fernando Camilo, personal communication).  
Source 2 is highly variable and has a
steep spectrum -- over the course of the ATCA observations
the flux density of this source varied by more than a factor of two,  and its
average flux densities at 1408 and 1708 MHz are $714 \pm 30 \mu$Jy and
$454 \pm 50 \mu$Jy respectively, corresponding to 
a spectral index
of $\alpha = -2.3 \pm 0.6$.  Therefore, it is highly likely
that Source 2 is 47TucJ.

EDITOR: INSERT FIGURE 5 HERE

Based on number counts of faint field radio sources, we
would expect, by chance, to detect about $\sim 0.7$ sources brighter
than $200 \mu$Jy within our search radius \cite{con84}.    
Thus there is no strong statistical reason to believe that
either of sources 1 or 3 is a pulsar.
Additionally, we find no evidence
of extended diffuse emission in the cluster.   After convolution of the image
with both a $25''$ and a $40''$ beam we find $3-\sigma$
limits on emission from the
center of the cluster of 0.9 and 1.5 mJy per beam respectively.  Although
these beams are large enough to encompass the most recent measurement
of the core radius of 47 Tucanae, $12\farcs2$ \cite{dpsg+96}, we note that 
none of the identified pulsars lie within $45''$ of the cluster center.
If the observed pulsars, rather than the reported core radius, provide
the best estimate of the distribution of fainter pulsars in the cluster,
than one would need perhaps as many as ten beams to properly cover the cluster,
and the limit on extended emission would be, at a minimum, several
mJy.

EDITOR: INSERT TABLE 3 HERE

\section{Discussion}

Our observations of Terzan~5, Liller~1 and NGC~6544 have revealed 
steep-spectrum 
radio emission even though these clusters do not possess known
pulsars (or, in the case of Terzan~5, known pulsars coincident
with the sources of most of the observed radiation).
In this section we show that this
emission is almost certainly produced by a significant population of pulsars
so far undetected by pulsed searches.

We have therefore used a Monte Carlo
simulation to determine 
whether our observations of the total radio emission
of Terzan~5 are consistent with the standard luminosity
function (as shown in Equation 1).   
Pulsar populations were created to simulate
Terzan~5 using a range of 
$L_{min}$, $p$ (the power-law exponent, see Equation 1), 
and total pulsar population.  For each
set of parameters, we determined the probability of creating a population
whose brightest two pulsars were within a factor of two of
the two brightest point sources in our Terzan~5 image (the 
eclipsing pulsar and the steep spectrum source north of the
the cluster center) and whose remaining total luminosity agreed,
again within a factor of two, with the 
observed total luminosity of the central region of the cluster.
As was found in FG90, the ability to fit the data is largely
independent of the assumed $L_{min}$.  However, the quality of our
fits depends strongly on
$p$.  The 90\% confidence interval 
is found to be $1.6 < p < 2.4$, with the
best fit at $p = 1.85$.  
This striking agreement with the well-observed
field luminosity function and the M15 pulsar luminosity function
derived by Anderson (1992) \nocite{and92}
makes it appear highly probable  that all
of the radio emission observed from Terzan~5 is produced by pulsars.

Our observations of Liller~1 and NGC~6544 are, however, significantly
less informative.  
All of our detections of Liller~1 are with 
beamsizes {\it larger} than the cluster core
radius.  Thus our inability to resolve the emission and study its
luminosity function is not surprising.  Nonetheless the signal-to-noise 
ratios of our detections are good enough to allow us to obtain an
accurate radio position in spite of our large beam.   In Table~4,
 we compare the position of the radio emission with that
of the optical cluster and the X-ray source in Liller~1.  All radio
positions agree; therefore we have reported only the 6~cm position, which
has the smallest error ellipse. 

EDITOR: INSERT TABLE 4 HERE

Although the reported
position of the rapid X-ray burster in Liller~1 \cite{ghsm+84} is 
offset from the radio position by about 8 arcseconds, or about
$4.5 \sigma$, evidence has recently been growing of a correlation
between the 6 cm flux density of Liller~1 and the X-ray strength
of MXB~1730-335 \cite{gfl+98,fgl+98,rmf+98}.    
The agreement between the radio source and the position 
of the center of the globular cluster reported by Liller \cite{lil77} is good;
however, the  
position of the optical center of the cluster is 
controversial.   Grindlay \etal (1984) report a position which disagrees
by several arcseconds
with that of  Liller.
However, Grindlay \etal used U band plates for their work 
whereas Liller worked in the R and I bands.  As the visual extinction
to this object is estimated at $\sim10$ magnitudes  \cite{dj93},
we have used Liller's position in the table.

Although the astrometric situation remains uncertain, an identification
of our 6 cm source with MXB~1730-335 could help explain the radio
spectrum of Liller 1.  Figure~6 shows a plot of the
radio data versus observing frequency. 
Pulsar spectra are usually steeper
at shorter rather than longer wavelengths \cite{mt77}.  
The object in Liller~1, however, shows the opposite trend.  
Indeed, 
our 20 cm flux density is $100\,\mu$Jy greater than the $4\sigma$ A
configuration limit of
Johnston, Kulkarni and Goss (1991).  Averaging our 20 cm data with
theirs would make this spectral break even more evident.  
We think it probable then that we may indeed be seeing flux from
the Liller~1 X-ray source in our 6 cm data; given
that our 20 and 90 cm observations
have flux densities significantly brighter than
the 6 cm flux densities observed at X-ray outburst (the 90 cm
flux density is more than a factor of $\sim 40$ brighter than outburst
levels),
we doubt this 
object
contributes significantly to our 90 or 20 cm images.    Indeed,
observations of X-ray binaries have typically found centimeter
radio spectra that are inverted or flat, and rather than steeply 
declining as seen here \cite{hje88}.

EDITOR: INSERT FIGURE 6 HERE


Given that our Terzan~5 observations confirm that the standard field
pulsar luminosity function can be applied to cluster millisecond
pulsars, we can use our Monte Carlo simulations to
estimate the number of pulsars in these clusters.
Assuming that the minimum 20 cm pulsar luminosity is $0.3$~mJy~kpc$^2$
(see Fruchter and Goss 1990 for a discussion of this assumption),
we estimate that there are between 60 and 200 pulsars in Terzan~5.
Similarly using Equation~2, 
we can use the total
luminosity of Liller~1 to estimate a population
of $\sim 15$ pulsars. 
A similar number is indicated
for the other cluster with an apparent undetected pulsar
in our VLA sample, NGC 6544. 
While these estimates of total numbers depend upon the assumed
(an not well known)
minimum pulsar luminosity, $L_{min}$,  it should be
noted that the relative expected number of pulsars in different clusters
is independent of the assumed $L_{min}$.     Furthermore, the absolute
numbers derived from this method will in general underestimate the
total number of pulsars.
As noted earlier, all of the identified sources
in 47 Tuc lie outside of a core radius from the center of the cluster,
and similarly the bright sources in Terzan~5 are more than a core
radius from the cluster center.  This, however, is not 
entirely unexpected:  a large
fraction of millisecond pulsars are thought to be produced as
a result of a collision of a binary with a third star (quite possibly
the neutron star) and this interaction can scatter the binary into
the outskirts (or indeed out) of the binary. Depending on the
mass distribution of the cluster, this may produce a significant
enhancement of the pulsar density at large radii \cite{sp95}.  
Thus,
the measurement of the diffuse radio emission within
a core radius of the cluster
will underestimate the number of weak pulsars
in the cluster.

Yet, if the radio spectra and luminosities of these
sources are consistent
with their being pulsars, why then have a substantial
fraction of the bright sources been missed 
by pulsed searches?
There are two obvious possibilities:  1) Binary
pulsars may be lost to pulsed surveys due to doppler smearing, even
though their radio emission remains detectable by
interferometric observations \cite{fg90,knr90,jkg91};
2) The pulses may be smeared beyond detectability by
interstellar dispersion and scattering.
Indeed, all three of the clusters found in this survey to
have substantial emission
unassociated with know pulsars
are located in the direction of the Galactic center
at uncertain but great ($\sim 10$ kpc) distances.  The resulting
expected dispersion measures are large.  Indeed, the
11 ms pulsar
in Terzan~5, PSR~1744-24A, has a dispersion measure of 230 cm$^{-3}$ pc, making
searches for pulsars with periods close to 1 ms extremely difficult.   
Searches are presently underway at Parkes which incorporate
software capable of detecting accelerated pulsars, or which
coherently de-disperse the radio signal,  thus largely eliminating
the effect of dispersion (although not scattering) on the pulsed signal.
One may hope that these searches will 
find the pulsed signals associated with the pulsars apparent in
our data.

Yet, one of the largest puzzles of these observations may come from
the clusters where pulsars were detected.  Our results imply that the flux
density of NGC~6440, NGC~6539 and NGC~6624 are all dominated by a single
pulsar.  In FG90 we used the total integrated
flux density of pulsars in globular clusters to estimate the total
number of pulsars in the Galactic globular cluster system.   We estimated
the number to lie between 500 and 2000.  The fact that so many of our
clusters are dominated by a single pulsar would tend to push the best
estimate down to the lower end of this range.  Thus 
our latest result only compounds a problem pointed out in FG90.
While 10\% of Galactic LMXBs are in clusters, only 1\% of their proposed
descendants live there.    Is this, as we noted in FG90, related to
the apparent lower luminosity of globular LMXBs -- do they spin up their
pulsars more slowly?  Or, is it just that objects in binaries are occasionally
expelled from the clusters (either with or without their companion, depending
upon the interaction) and as the lifetime of a MSP is longer than that
of an LMXB, there is a greater chance that it will be expelled during
its lifetime?  Or are the LMXB's formed by collision in clusters 
somehow less fecund than the native born binaries of the disk?  
In these discrepant 
statistics may lie an important clue to understanding
the evolution of cluster binaries and the formation of millisecond pulsars.

\section{Acknowledgements}

We would like to thank Tim Cornwell, whose programs and assistance 
were invaluable to our 90 cm imaging, and David McConnell, Shri Kulkarni
and Helen Johnston for supporting
our use of their archival data.  
Taft Armandroff graciously provided an I-band image of Terzan~5 obtained
at the CTIO 1-m.  We also thank an anonymous referee for a very careful
reading of our manuscript and many helpful comments.
The radio observations presented
here were done at the Very Large Array of the National 
Radio Astronomy
Observatory,
which is  a facility of
the National Science Foundation operated under cooperative
agreement by Associated Universities , Inc. and at the Australia Telescope,
which is funded by the Commonwealth of Australia for 
operation as a national facility by CSIRO.   

\eject


\eject

\begin{table}[h]
\begin{center}
{TABLE 1 \\
{\sc Terzan 5 Source Positions and Flux Densities}}
\vskip 0.2 cm
\begin{tabular}{l l l l l r} \hline \hline
{Source} & \multicolumn{1}{c}{$\alpha_{1950}$} & {error} &
\multicolumn{1}{c}{$\delta_{1950}$} & {error} &{Flux Density}$^1$\\
& & & & &  {\tenrm ($\mu$Jy)} \\ \hline
N'& \RA 17 45 00 50  & \erra 0 09 & \dec -24 45 21 1 & \errd 0 8 & 7500 \\
N & \RA    17 44 59 985 & \erra 0 005 &\dec -24 45 35 58 & \errd 0 05 & 1420 \\
A (PSR 1744-24A) & \RA 17 44 57 710 & \erra 0 005 &\dec -24 45 38 20 & \errd 0 05 & 1340  \\
C & \RA 17 45 00 41 & \erra 0 012 &\dec -24 45 46 0 & \errd 0 18 & 230 \\
S & \RA 17 45 00 13 & \erra 0 015 &\dec -24 45 51 8 & \errd 0 20 & 230 \\ \hline
\end{tabular}
\end{center}
$^1$ Flux densities are given per beam.  All flux densities, with the exception
of that for the source N' are for the $2\farcs9$ beam of the 20 cm image
shown in Figure~1. These sources have an error of 28~$\mu$Jy/beam.  
Source N' is 
only visible in the 90 cm image, and the flux density given is for that image 
(Figure~4).  The 90 cm flux density error is 1.3~mJy/beam.
\end{table}

\eject

\begin{table}[h]
\begin{center}
{TABLE 2 \\
{\sc Sources in Clusters Dominated by Known Pulsars}}
\vskip 0.2 cm
\begin{tabular}{l l l l l} \hline \hline
{Cluster} & {Band} & \multicolumn{1}{c}{$\alpha_{2000}$} & \multicolumn{1}{c}{$\delta_{2000}$}  & Flux Density\\  
  & & & &  {\tenrm ($\mu$Jy)} \\ \hline
NGC~6539 & 20 cm & $18^h\,04^m\,49^s.87 \pm 0.01$ & $-07^\circ\,35'\,24\farcs7 \pm 0.1 $&  $485 \pm 85$ \\
NGC~6539 & 6 cm &  &  & $< 100 $ \\ 
NGC~6539 & Pulsar$^1$ & $18^h\,04^m\,49^s.89 \pm 0.01$ & $-07^\circ\,35'\,24\farcs7 \pm 0.1 $ & $500 \pm 200$  \\ 
NGC~6539 & Optical$^2$  & $18^h\,04^m\,49^s.72 $ & $-07^\circ\,35'\,09\farcs1  $  & \\ \hline
NGC~6624 & 20 cm & $18^h\,23^m\,40^s.46 \pm 0.01$ & $-30^\circ\,21'\,39\farcs6 \pm 0.1 $ & $420 \pm 45$  \\
NGC~6624 & 6 cm &  &  & $ 116 \pm 26$ \\
NGC~6624 & Pulsar$^3$ & $18^h\,23^m\,41^s.46 \pm 0.02$ & -$30^\circ\,21'\,40\farcs1 \pm 0.4 $ & $400 \pm 100$  \\
NGC~6624 & Optical$^2$ & $18^h\,23^m\,40^s.66$ & -$30^\circ\,21'\,38\farcs8  $ &   \\ \hline
NGC~6440 & 20 cm & $17^h\,48^m\,52^s.66 \pm 0.01$ & $-20^\circ\,21'\,39.\farcs3 \pm 0.1 $ & $750 \pm 100 $ \\
NGC~6440 & 6 cm & & & $90 \pm 30$ \\ 
NGC~6440 & Pulsar$^4$ & $17^h\,48^m\,52^s.6 \pm 0.1 $ & $-20^\circ\,22'\,30 \pm 30$ & $1500 \pm 500$ \\
NGC~6440 & Optical$^2$ & $17^h\,48^m\,52^s.65  $ & $-20^\circ\,21'\,34.\farcs5 $ & \\ \hline
\end{tabular}
\end{center}
$^1$D'Amico et al.~(1993) \\
$^2$Shawl and White (1986) \\
$^3$Biggs et al.~(1994) \\
$^4$Manchester et al.~(1989) 

\end{table}

\eject
 
\begin{table}[h] 
\begin{center}
{TABLE 3 \\
{\sc Positions and Flux Densities of 47 Tucanae Sources}}
\vskip 0.2cm
\begin{tabular}{ l  l l r } \hline\hline
{Source} & \multicolumn{1}{c}{$\alpha_{2000}$} &
\multicolumn{1}{c}{$\delta_{2000}$}  & {Flux Density} \\
  & & & {\tenrm (mJy)} \\ \hline
C & $00^h\,23^m\,50^s.34 \pm 0.07$ & $-72^\circ\,04'\,31\farcs5 \pm 0.3$ & $0.33 \pm 0.03 $ \\
D & $00^h\,24^m\,13^s.78 \pm 0.04$ & $-72^\circ\,04'\,44\farcs9 \pm 0.3$ & $0.40 \pm 0.05$ \\
1 & $00^h\,23^m\,40^s.80 \pm 0.03$ & $-72^\circ\,05'\,12\farcs.2 \pm 0.1$ & $0.78 \pm 0.03 $ \\
2 & $00^h\,23^m\,59^s.33 \pm 0.03$ & $-72^\circ\,03'\,59\farcs1 \pm 0.2$ & $0.65 \pm 0.03 $ \\ 
3 & $00^h\,24^m\,13^s.36 \pm 0.07$ & $-72^\circ\,03'\,32\farcs6 \pm 0.4$ & $0.30 \pm 0.03$ \\ \hline
\end{tabular}
\end{center}
\end{table}

\eject

\begin{table}[h]
\begin{center}
{TABLE 4 \\
{\sc Liller 1 Source Positions}}
\vskip 0.2 cm
\begin{tabular}{l r r r r } \hline \hline
{Band} & \multicolumn{1}{c}{$\alpha_{2000}$} & {error} & \multicolumn{1}{c}{$\delta_{2000}$} & {error} \\  \hline
Radio &\RA      17 33 24 56 & \erra 0 03 & \dec -33 23 19 8     & \errd 0 5 \\
Optical$^1$ &\RA    17 33 24 47 & \erra 0 08  & \dec -33 23 20 2    & \errd 1 0 \\
X-ray$^2$ &\RA      17 33 24 09 & \erra 0 08 & \dec -33 23 16 4     & \errd 1 0 \\ \hline
\end{tabular}
\end{center}
$^1$Liller~(1977) \\
$^2$Grindlay et al.~(1984)
\end{table}

\eject

\section{Figures}

{\bf Figure 1:} Radio contours are shown superposed on an I band
image of the rich cluster Terzan 5.   
The 20 cm radio image combines data taken several different
VLA configurations and has a beamsize of $2\farcs9 \times 2\farcs9''$. 
Contour levels are -100, 100, 200, 300, 400, 500 and
1000 $\mu$Jy/beam.   The r.m.s. noise level in the image is 28~$\mu$Jy.

{\bf Figure 2:} 20 cm image of Terzan~5 using a larger
synthesized beam (robust weight $R=1$) than in Figure~1.  This
image shows the large extent of the diffuse radio emission
centered on the core of the globular cluster.  The beam is 
$6\farcs3 \times 4\farcs4$ and the image has an r.m.s. noise
of 23~$\mu$Jy/beam. the Contour levels are -100, 100, 200, 
300, 400, 750, 1000 and 2000 $\mu$Jy/beam.

{\bf Figure 3:} 6 cm radio contours (beamsize $8'' \times 5''$, PA=$6^\circ$) are
are shown superposed on a 20 cm image (greyscale) of
the cluster Terzan~5.  The r.m.s. noise of the image
is 25~$\mu$Jy/beam; the radio contours are  -50, 50, 80, 100, 120, 140 and
160 $\mu$Jy/beam.

{\bf Figure 4:} 90 cm radio contours (beamsize $18'' \times 12''$, PA=$67^\circ$
are shown superposed on a 20 cm image of the cluster Terzan~5
(Figure~1, now shown in greyscale).
While the central emission displays a
steep spectral index characteristic of pulsars, the one identified
pulsar in the field, PSR~1744-24A is not detected, even though it
was strong a strong source in the simultaneous 20 cm observations.  
This pulsar appears to be continuously eclipsed at 90 cm.
The source N' is undetected in either the 20 or 6 cm images
of this cluster, implying a 90 to 20 cm spectral index, $\alpha, > 2.5$.
The radio contours are $-6$, $-3$, 3, 6, 12, 24 and 48~mJy/beam; the
r.m.s. noise of the image is $1.3$~mJy/beam;.

{\bf Figure 5:}  The ATCA image of the rich globular cluster 47 Tucanae.
The center of the cluster is marked with an asterix, and the
radio timing positions of pulsars 47TucC and 47TucD are marked
with crosses.  Two of the sources are identified, based on their
positions, as being pulsars C and D from this cluster, and,
as discussed in the text, Source 2 is most likely pulsar 47TucJ.  
Sources 1 and 3 remain unidentified.
The observations were obtained at a wavelength of
approximately 20 cm, and the beam is nearly circular
with a $6''$ FWHM.  The contours shown 
are -120, 120, 240, and 480~$\mu$Jy/beam; the r.m.s. noise of the image
is 32~$\mu$Jy/beam.

{\bf Figure 6:} The flux density of the radio emission from Liller~1 as
a function of observing frequency.  As discussed in the text, the
flux densities at 330 MHz (90 cm) and 1400 MHz (20 cm) are most
likely dominated by emission from one or more pulsars, while
some or most of the 4860 MHz (6 cm) emission may arise from
an X-ray binary in the cluster.  The line connecting the 
90 and 6 cm flux densities is used only to guide the eye,
and emphasizes the particularly sharp drop in flux density between
90 and 20 cm.

\eject

\begin{figure}[h]
\centerline{\psfig{file=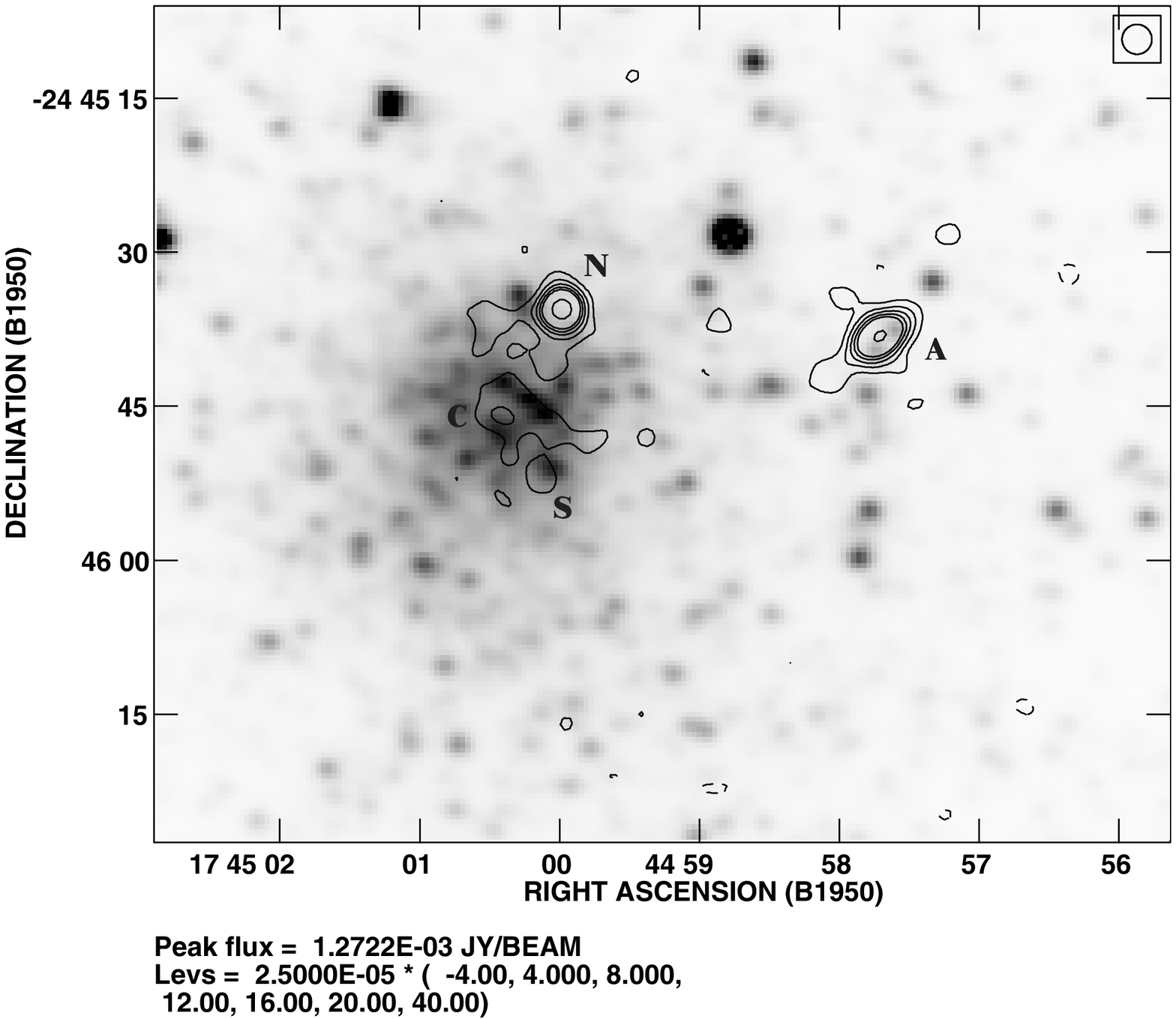,width=5.5in}}
\caption[]{}
\end{figure}
\eject

\begin{figure}[h]
\centerline{\psfig{file=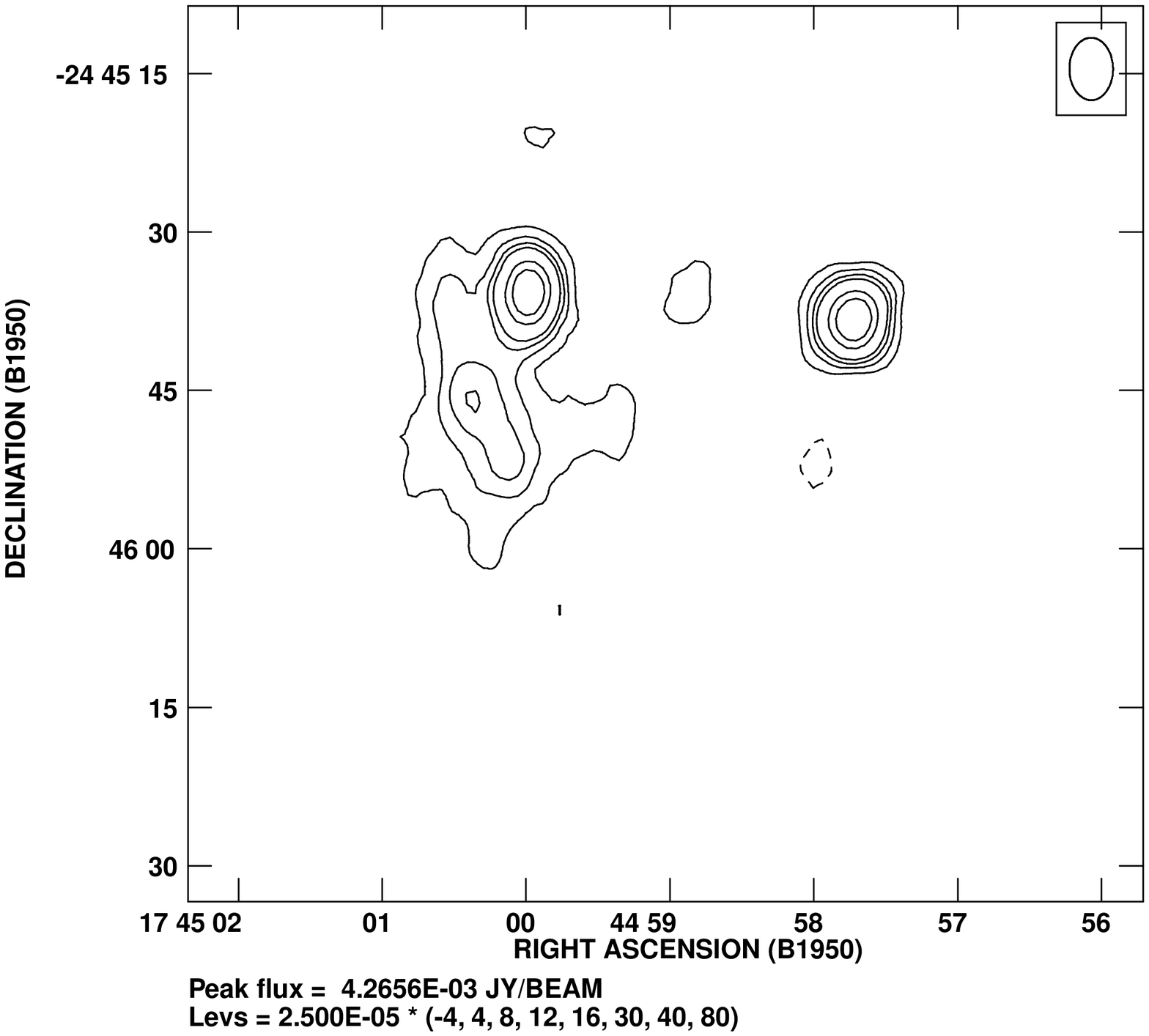,width=5.5in}}
\caption[]{}
\end{figure}

\eject

\begin{figure}[h]
\centerline{\psfig{file=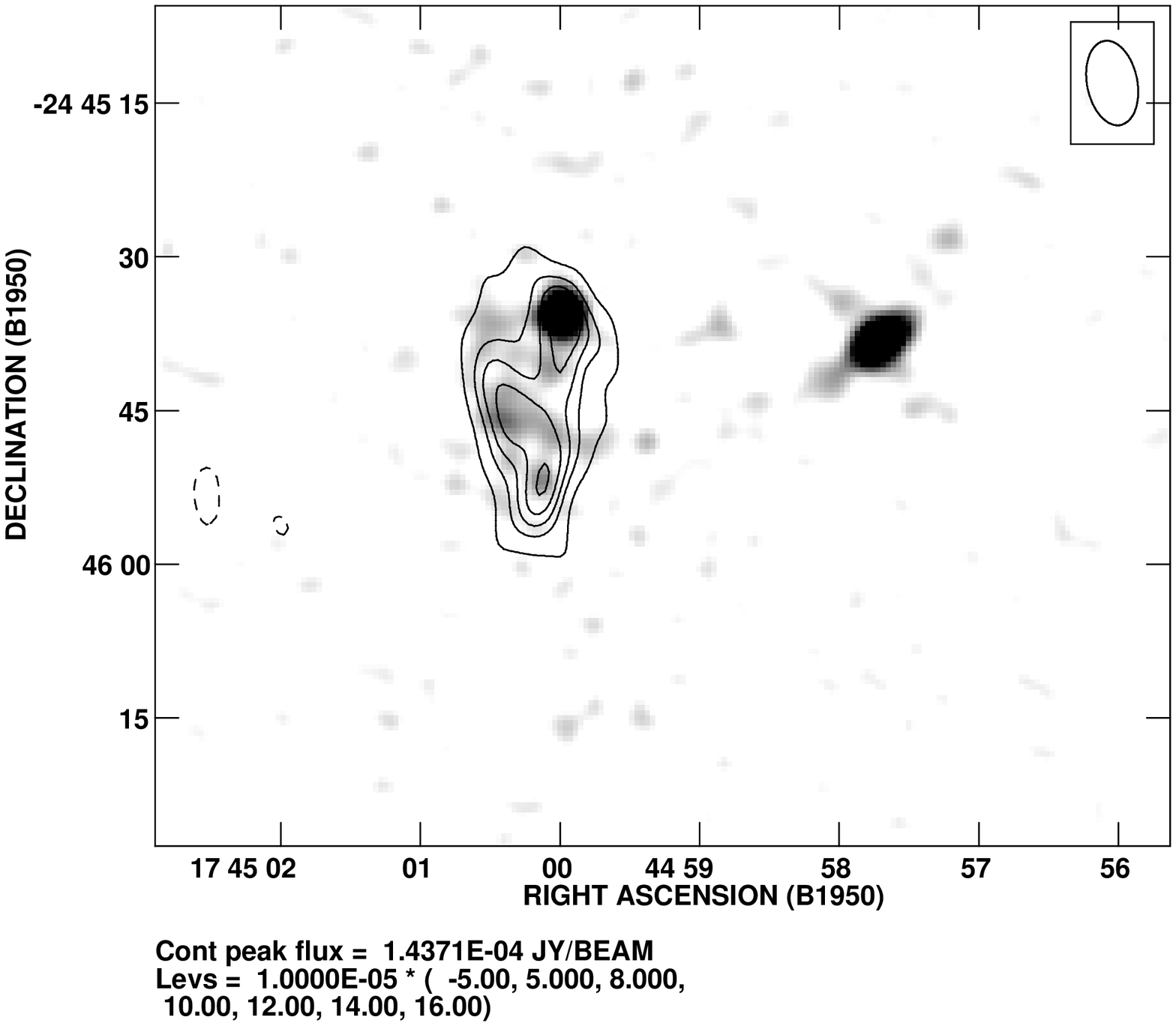,width=5.5in,angle=-90}}
\caption[]{}
\end{figure}

\eject

\begin{figure}[h]
\centerline{\psfig{file=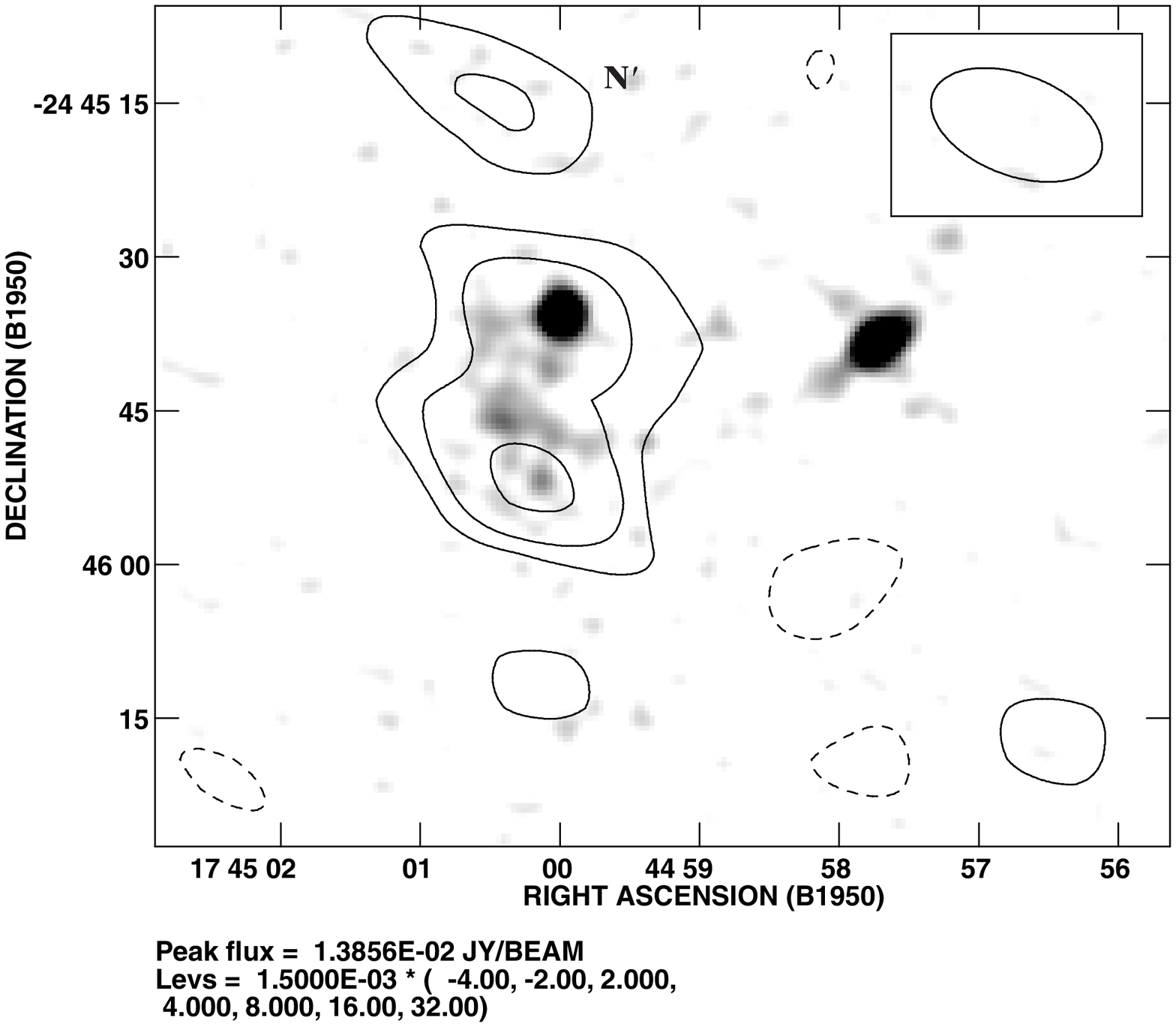,width=5.5in}}
\caption[]{}
\end{figure}

\eject

\begin{figure}[h]
\centerline{\psfig{file=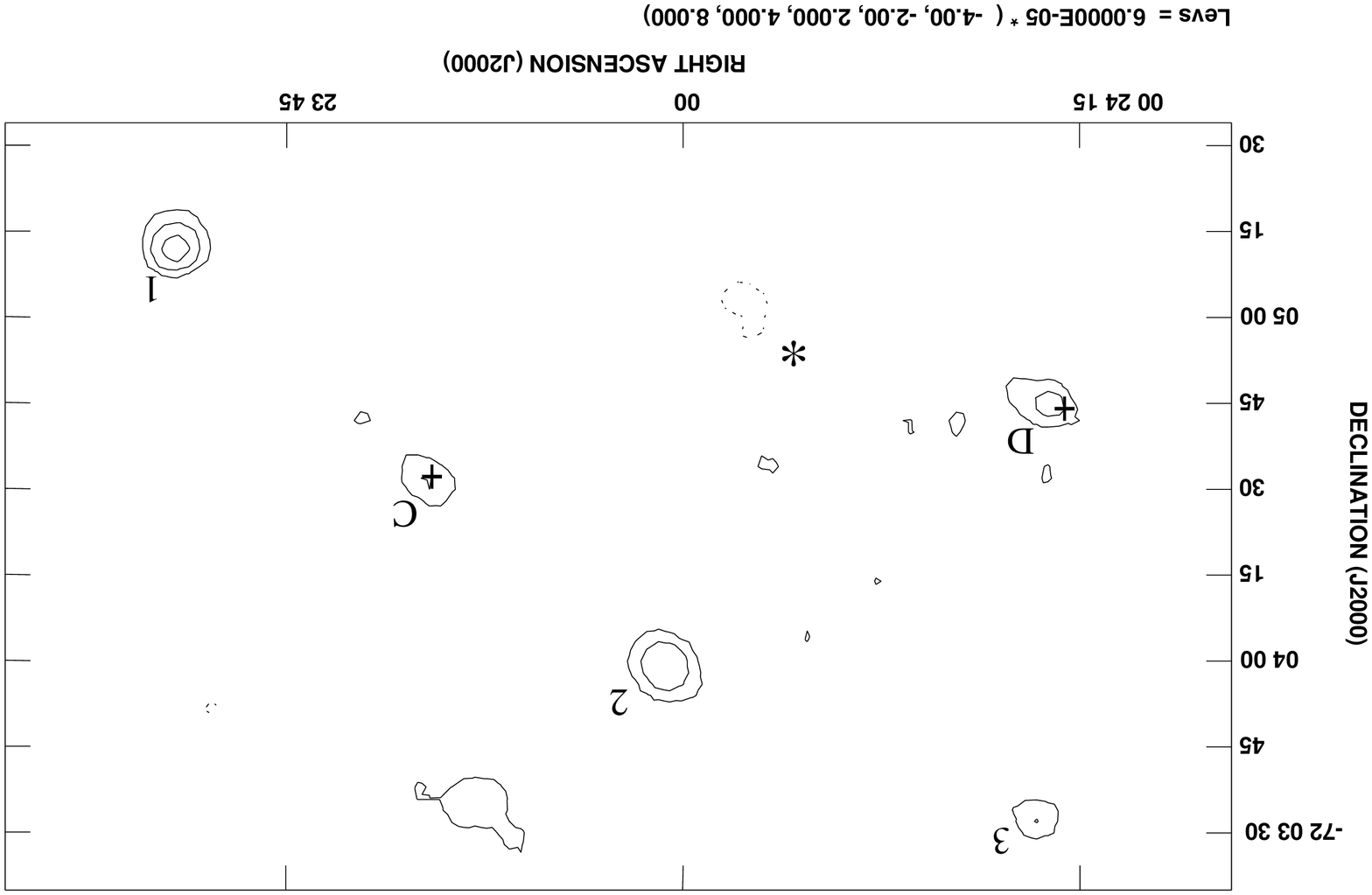,width=6in,angle=-90}}
\caption[]{}
\end{figure}

\eject

\begin{figure}[h]
\centerline{\psfig{file=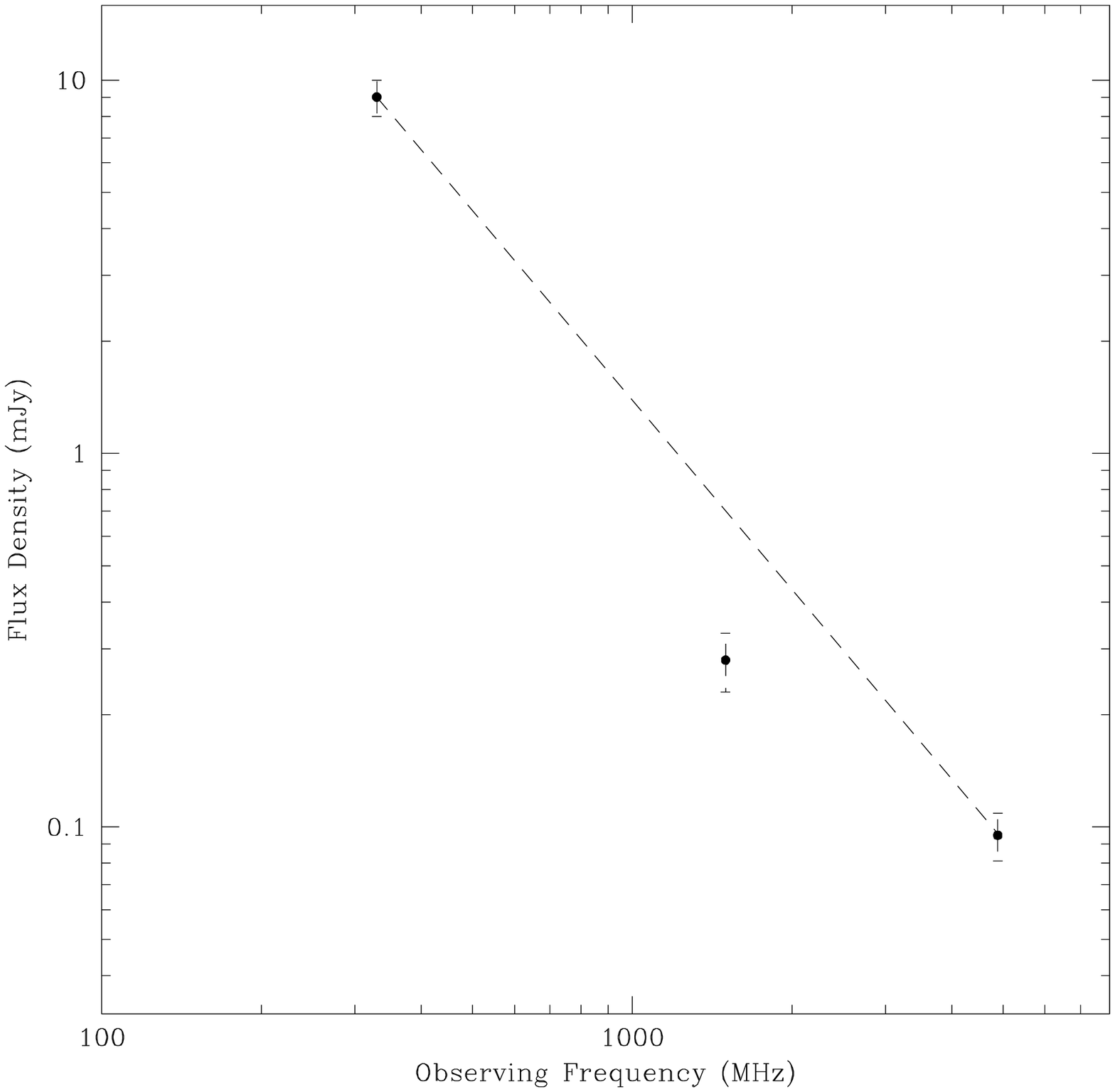,width=5.5in}}
\label{f:liller}
\caption[]{}
\end{figure}

\end{document}